\documentclass[aps,prl,twocolumn,showpacs,superscriptaddress]{revtex4}
\usepackage[T1]{fontenc}
\usepackage[latin1]{inputenc}
\usepackage{graphicx}
\usepackage{amssymb}



\begin{document}

\title{Ground state cooling of nanomechanical resonator via parametric linear coupling}

\author{L. Tian}

\affiliation{School of Natural Sciences, P. O. Box 2039, University of California,
Merced, CA 95344}


\affiliation{Department of Applied Physics and Edward L. Ginzton Laboratory, Stanford
University, Stanford, CA 94305}

\date{\today}

\begin{abstract}
We present a ground state cooling scheme for a nanomechanical resonator linearly
coupled with a LC oscillator. The linear coupling, when periodically modulated at red detuning, up-converts the low-frequency nanomechanical mode to the high-frequency LC oscillator mode and generates backaction force that can cool the resonator to its ground state in the resolved-sideband regime. We also study the effect of the quantum backaction noise on the cooling due to the counter rotating term in the linear coupling.  The scheme can be compared with laser cooling for the atomic systems and can be realized in superconducting circuits.
\end{abstract}
\maketitle

Nanomechanical systems with ultra-high quality factor (Q-factor) have been demonstrated to approach the quantum limit  in recent experiments \cite{SchwabPToday2005,HarrisNature2008,Kippenberg2008,RegalNPhys2008}.  The study of the quantum behavior in such systems can have profound impact on various topics including the detection of weak forces \cite{BraginskyScience1980}, the
study of classical-and-quantum boundary in macroscopic objects \cite{LeggettPRL1985,LFWeiPRL2006}, and quantum entanglement and quantum information \cite{MarshallPRL2003}. Couplings between nanomechanical systems with  solid-state devices \cite{BlencowePRep2004} or atomic systems \cite{TianZollerPRL2004} have been widely studied and can facilitate the implementation of quantum control and quantum engineering in such systems \cite{JacobsPRL2007_2}.

Ground state cooling is crucial for the quantum engineering of nanomechanical systems \cite{KnobelNature2003}. Recently, cooling of nanomechanical modes from room temperature to a few Kelvin has been achieved via dynamic backaction force or active feedback cooling in optical cavities and solid-state circuits \cite{SchwabScience2006,FlowersJacobsPRL2007,BrownPRL2007,Karrai2004}.  
Furthermore, the resolved-sideband regime has been reached in experiments, which is a key step in the preparation of the nanomechanical modes to the ground state \cite{Kippenberg2008,RegalNPhys2008}. In theory, full quantum mechanical treatments have been developed to study ground state cooling of the nanomechanical modes via radiation pressure force \cite{WilsonRaePRL2007,GirvinPRL2007,GenesPRA2008,ClerkPRL2006}. It has also been proposed that cooling can be achieved by coupling the nanomechanical modes to quantum two-level systems (qubits) in solid-state devices \cite{MartinPRB2004,WilsonRaePRL2004}.

Solid-state electronic circuits, with their flexibility in device layout and parameter engineering, provide diverse forms of coupling between nanomechanical modes and other degrees of freedom in the circuit. The radiation pressure-like force between a nanomechanical mode and a superconducting resonator has been explored for the cooling of the nanomechanical mode \cite{RegalNPhys2008,BrownPRL2007}.  In this work, we study a novel ground state cooling scheme where the nanomechanical resonator couples \emph{linearly} with a LC oscillator mode \cite{TianResonators}.  For constant coupling, the LC oscillator together with its thermal bath can be viewed as a structured-thermal bath that is in equilibrium with the nanomechanical mode \cite{Grifoni2004},  with no cooling. However, when the coupling is periodically modulated at red detuning, the low-frequency nanomechanical quanta are up-converted to the high-frequency LC oscillator quanta which are subsequently dissipated in the circuit, and cooling can be achieved.  Here, we study the cooling process with the input-output theory in quantum optics \cite{WilsonRaePRL2007,GirvinPRL2007,WallsBook1994} and show that the nanomechanical mode can reach the ground state in the resolved-sideband regime. The stationary occupation number of the nanomechanical mode (final phonon number) is limited by the quantum backaction noise which is due to the counter rotating term in the Bogoliubov linear coupling \cite{KippenbergPreprint}.  A close analogue between this scheme and the laser cooling for atomic systems can be drawn \cite{WinelandPRA1987}. The cooling scheme can also be explained by a semiclassical circuit theory where the dynamical backaction force acts as a frictional force on the nanomechanical vibration. The scheme can be implemented by coupling the nanomechanical mode capacitively with a superconducting resonator mode \cite{MakhlinRMP2001,HouckNature2007} and using an external gate voltage to control the coupling. Note that periodically modulated linear coupling was studied previously for generating entanglement and squeezed state in nanomechanical modes \cite{TianResonators}, which is a central element in implementing quantum protocols such as quantum teleportation.

\begin{figure}
\includegraphics[bb=81bp 360bp 540bp 618bp,clip,width=7.5cm]{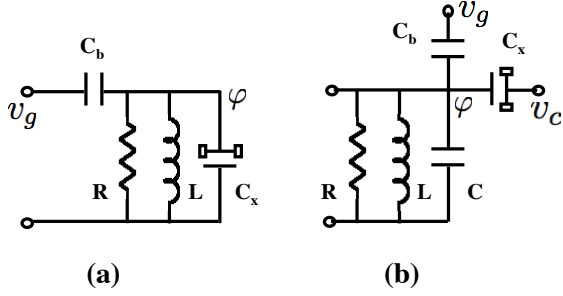} 
\caption{Circuits of nanomechanical resonator capacitively coupling with a LC oscillator. (a) radiation pressure-like coupling and (b) linear coupling modulated by gate voltage. }
\label{fig1} 
\end{figure}

Consider a nanomechanical resonator capacitively coupling with a LC oscillator in a solid-state circuit. The nanomechanical resonator forms one plate of the coupling capacitor, where the capacitance can be expressed as $C(x)=C_{x0}(1-x/d_{0})$ in terms of the vibrational displacement $x$ of the nanomechanical mode and the distance $d_{0}$ between the capacitor plates.  Two schematic  circuits are presented in Fig.~\ref{fig1}, which give two types of coupling. The total Hamiltonian has the general form $H_{t}=\hbar\omega_{a}a^{\dagger}a+H_{c}$ with $\omega_{a}$ being the frequency and $a$ ($a^{\dagger}$) being the annihilation (creation) operator of nanomechanical mode, and $H_{c}$ being the Hamiltonian of the LC oscillator.  In circuit (a), $H_c$ includes a capacitive energy $p_{\varphi}^{2}/2C(x)$ and an inductive energy $\varphi^{2}/2L$, where $\varphi$ is the phase variable labeled in the circuit and $p_{\varphi}$ is the conjugate momentum. And $H_{c}$ can be written as
\begin{equation}
H_{c}=\hbar\omega_{b}b^{\dagger}b-\frac{\hbar\omega_{b}}{2}\frac{\delta x_{0}}{d_{0}}(a+a^\dagger) b^{\dagger}b \label{Ha}
\end{equation}
when expanded to the first order of $x$. Here, $\omega_{b}=(LC_{x0})^{-1/2}$ is the frequency of the LC oscillator, $b$ ($b^{\dagger}$) is the annihilation (creation) operator of the phase variable, and $\delta x_0$ is the quantum displacement of the nanomechanical mode with $x=\delta x_0 (a+a^{\dagger})$. The coupling between the nanomechanical mode and the LC oscillator in this circuit is hence a radiation pressure-like coupling \cite{BrownPRL2007}. It was shown that resolved-sideband regime for the nanomechanical mode can be reached in this circuit \cite{RegalNPhys2008}. 

Below we focus on circuit (b), where the nanomechanical resonator couples with  the phase $\varphi$ via an applied gate voltage $v_{c}$. The Hamiltonian of the LC oscillator is 
\begin{equation}
H_{c}=\frac{(p_{\varphi}-C_{x0}v_{c}x/d_{0})^{2}}{2(C_{\Sigma0}-C_{x0}x/d_{0})}+\frac{\varphi^{2}}{2L},\end{equation}
with the total capacitance $C_{\Sigma0}=C+C_{g}+C_{x0}$. Note that the gate voltage $v_{g}=-v_{c}C_{x0}/C_{b}$ is applied to the capacitance $C_{b}$ to balance the linear driving on the momentum $p_{\varphi}$ due to $v_{c}$. It can be derived that
\begin{equation}
H_{c}=\hbar\omega_{b}b^{\dagger}b +g_{r} (a+a^\dagger)b^{\dagger}b - ig_{l}(a+a^\dagger)(b-b^{\dagger}) 
\label{Hb}
\end{equation}
which includes both a radiation pressure-like coupling with the coupling constant
\begin{equation}
g_{r}=-\frac{\hbar\omega_{b}}{2}\frac{C_{x0}}{C_{\Sigma0}}\frac{\delta x_{0}}{d_{0}}\label{gr}
\end{equation}
and a Bogoliubov linear coupling with the coupling constant 
\begin{equation}
g_{l}=-iC_{x0}v_{c}\sqrt{\frac{\hbar \omega_b}{2 C_{\Sigma0}}}\frac{\delta x_0}{d_{0}}\label{gl}
\end{equation}
that depends on the applied voltage $v_{c}$. The frequency of the LC oscillator is now $\omega_{b}=(LC_{\Sigma0})^{-1/2}$.

We choose the following parameters which are realizable in superconducting circuits: $C_{x0}=0.6\,\textrm{fF}$, $C_{\Sigma0}=2.5\,\textrm{fF}$, $\omega_{b}=7.5\,\textrm{GHz}$, and $v_{c}=25\,\textrm{mV}$.
The ratio between the linear coupling and radiation pressure-like coupling can be derived to be $g_{l}/g_{r}=v_{c}\sqrt{2C_{\Sigma0}/\hbar\omega_{b}}$. With the above parameters, we have $g_{l}/g_{r}\approx 10^{3}$ and $g_{l}=2\,\textrm{MHz}$. In contrast to circuit (a), the linear coupling is the dominant term here. We will only consider the linear coupling in our discussion. With a constant driving voltage, Eq.~(\ref{Hb}) describes two harmonic oscillator modes in thermal equilibrium at the bath temperature $T_0$. However, the scenario changes when the coupling is periodically modulated by an RF voltage source $v_{c}(t)=2v_{c}\sin\omega_{d}t$ at a frequency $\omega_{d}$. In the rotating frame of the driving frequency, the total Hamiltonian becomes 
\begin{equation}
H_{t}^{rot}=\hbar\omega_{a}a^{\dagger}a-\hbar\Delta b^{\dagger}b+g_{l}(a+a^{\dagger})(b+b^{\dagger})\label{Ht_rot}
\end{equation}
with the detuning $-\Delta=\omega_b - \omega_d$ and the coupling $g_{l}$. Here, we study the red detuning regime with $-\Delta>0$.

Damping is a crucial factor in the cooling process. We assume the damping rate of the nanomechanical mode to be $\gamma_{0}$ and the damping rate of the LC oscillator to be $\kappa_{0}$, both of which are associated with a thermal bath at the temperature $T_{0}$ and the thermal occupation number $n_{i0}=(\exp(\hbar\omega_i/k_BT_0)-1)^{-1}$ for $i=a,b$. With $T_0=20\,\textrm{mK}$ and $\omega_{b}=7.5\,\textrm{GHz}$ for the LC oscillator, $n_{b0}\approx 10^{-8}$ and can be treated as zero. 

We use the input-output theory to study cooling in the linearly coupled system.  In the Heisenberg picture, the following operator equations can be derived
\begin{eqnarray}
\dot{a} & = & -i\omega_{a}a-ig_{l}(b+b^{\dagger})-\frac{\gamma_{0}}{2}a+\sqrt{\gamma_{0}}a_{in}\nonumber \\
\dot{b} & = & i\Delta b-ig_{l}(a+a^{\dagger})-\frac{\kappa_{0}}{2}b+\sqrt{\kappa_{0}}b_{in}\end{eqnarray}
as well as their conjugate equations. The operators $a_{in}$ and $b_{in}$ are noise operators for the corresponding thermal bath~\cite{WallsBook1994}. Similar to Ref. \cite{WilsonRaePRL2007,GirvinPRL2007}, the cooling rate can be derived as
\begin{equation}
\Gamma_{c}=\frac{4g_{l}^{2}\kappa_{0}|\Delta|\omega_{a}}{(\Delta^{2}-\omega_{a}^{2}+\frac{\kappa_{0}^{2}}{4})^{2}+\omega_{a}^{2}\kappa_{0}^{2}}\label{Ga_inout}
\end{equation}
in the weak coupling limit. And the stationary occupation number is 
\begin{equation}
n_{a}^{f}=\frac{\Gamma_{c}n_{0}+\gamma_{0}n_{a0}}{\Gamma_{c}+\gamma_{0}}\approx n_{0}+\frac{\gamma_{0}}{\Gamma_{c}}(n_{a0}-n_0)\label{naf_inout}
\end{equation}
when $\Gamma_{c}\gg \gamma_{0}$, and cooling of the nanomechanical mode can be achieved. Here, $n_{0}\approx \kappa_{0}^{2}/16\omega_{a}^{2}$ is the quantum backaction noise. Note when $-\Delta=\omega_a$, i.e. the driving frequency is at the first red sideband, we have
\begin{equation}
\Gamma_{c}=\frac{4g_{l}^{2}}{\kappa_{0}(1+\frac{\kappa_{0}^{2}}{16\omega_{a}^{2}})}.
\end{equation}
One interesting point is that a linear coupling of the form in Eq.~(\ref{Ht_rot}) can be derived from the radiation pressure-like coupling by linearizing the interaction 
relative to the amplitude of the driven cavity mode, which explains the similarity between the current result and that in the radiation pressure cooling schemes \cite{WilsonRaePRL2007,GirvinPRL2007,KippenbergPreprint}.

The periodical modulation of the linear coupling is the key to this cooling scheme. For constant gate voltage $v_{c}$, we can replace $-\Delta$ in Eq. (\ref{Ga_inout}) by $\omega_{b}$ and derive that $\Gamma_{c}\approx 4g_{l}^{2}\kappa_{0}\omega_{a}/\omega_{b}^{3}\ll \gamma_0$ with $\omega_{b}\gg\omega_{a}$ (even with a nanomechanical Q-factor of $10^{7}$). Then, $n_{a}^{f}\approx n_{a0}$ with no cooling. The two oscillator modes are hence in thermal equilibrium,  and the LC oscillator together with its bath acts as an effective thermal bath for the nanomechanical mode at the temperature $T_{0}$ \cite{Grifoni2004}. With periodical modulation of the coupling, the effective frequency of the LC oscillator becomes $-\Delta$ in Eq. (\ref{Ht_rot}), but its thermal occupation number is still $n_{b0}$. Hence, the LC oscillator can now be viewed as being at an effective temperature 
\begin{equation}
T_{eff}=T_{0}\frac{|\Delta|}{\omega_{b}}\ll T_{0}
\end{equation}
and acts as a cold reservoir that extracts energy from the nanomechanical mode. This is similar to the sympathetic cooling scheme where one ion (atom) species at a lower temperature is used to cool another ion (atom) species at a higher temperature via Coulomb coupling \cite{LarsonPRL1986}. The periodical modulation of the coupling up-converts the low-frequency nanomechanical quanta to the high-frequency LC oscillator quanta.

The above cooling scheme can also be viewed as a modified laser cooling scheme in the resolved-sideband regime where $\omega_{a} \gg\kappa_{0}$ \cite{WinelandPRA1987}.  With periodical modulation of the linear coupling at the red detuned frequency $-\Delta=\omega_a$, the term $a^{\dagger}b+b^{\dagger}a$ generates a resonant cooling transition with the transition rate (cooling rate) $A_{-}=4g_{l}^{2}/\kappa_{0}$. At the same time, the term $a^{\dagger}b^{\dagger}+ba$ generates an off-resonant heating transition with the transition rate (heating rate) $A_{+}\approx g_{l}^{2}\kappa_{0}/4\omega_{a}^{2}$, which is the origin of the quantum backaction noise. This analysis gives: $n_{0}=A_{+}/(A_{-}-A_{+})\approx \kappa_{0}^{2}/16\omega_{a}^{2}$, agreeing with our previous result. 

To clarify the role of the counter rotating term $a^{\dagger}b^{\dagger}+ba$, we study the cooling process by omitting this term in the linear coupling and solve the stationary occupation number of the nanomechanical mode with a master equation approach. We derive
 \begin{equation}
n_{a}^{f}=\left(1+\frac{4(\omega_{a}+\Delta)^{2}+\kappa_0^2}{4g_{l}^{2}}\right)\frac{\gamma_0}{\kappa_0}n_{a0}
\label{naf_fpe}
\end{equation}
to the first order of the damping rate $\gamma_{0}$. In contrast to Eq. (\ref{naf_inout}), 
the occupation number here is solely limited by the damping rate $\gamma_{0}$ of the nanomechanical mode. This result shows that the quantum backaction noise is a direct result of the counter rotating term. 

\begin{figure}
\includegraphics[bb=72bp 300bp 540bp 636bp,clip,width=7.5cm]{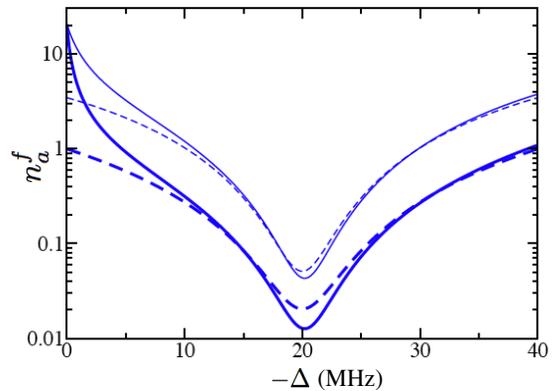} 
\caption{Occupation number $n_{a}^{f}$ versus detuning. Solid curves are
by Eq.~(\ref{naf_inout}) at $g_{l}=2\,\textrm{MHz}$ (thick line) and $g_{l}=1\,\textrm{MHz}$ (thin line). Dashed curves are by Eq.~(\ref{naf_fpe}) at $g_{l}=2\,\textrm{MHz}$ (thick line) and $g_{l}=1\,\textrm{MHz}$ (thin line). Other parameters are:  $\omega_{a}=20\,\textrm{MHz}$, $\gamma_{0}=2\,\textrm{KHz}$, and $\kappa_{0}=4\,\textrm{MHz}$.}
\label{fig2} 
\end{figure}

One experimental system to implement this scheme is a nanomechanical resonator capacitively coupling with a superconducting LC resonator \cite{HouckNature2007}. The damping rate of the LC oscillator can be tuned by using a dissipative element R in the circuit. With the progress in microfabrication, nanomechanical resonators with Q-factor exceeding $10^{7}$ can now be made \cite{HarrisNature2008} where $\Gamma_{c}\gg \gamma_{0}$. The occupation number $n_a^f$ is hence lower bounded by the quantum backaction noise $n_0$. In Fig.\ref{fig2}, we plot the stationary occupation number given by Eq.~(\ref{naf_inout}) with typical parameters in superconducting circuits. It is shown that cooling of the nanomechanical mode to $n_{a}^{f}=0.01$ can be achieved, starting from $T_0=20\,\textrm{mK}$ ($n_{a0}=20$). The results by Eq. (\ref{naf_fpe}) are also plotted for comparison. 

\begin{figure}
\includegraphics[bb=54bp 306bp 546bp 654bp,clip,width=7.5cm]{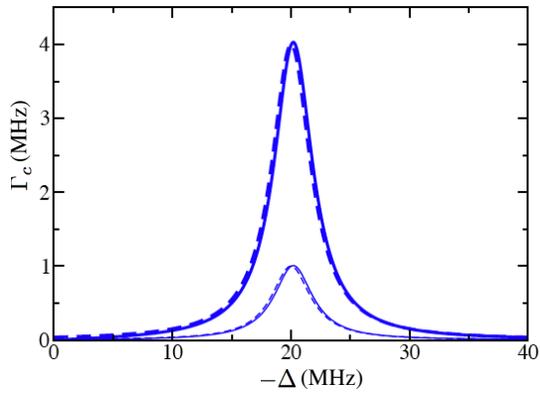} 
\caption{Cooling rate versus detuning. Solid curves are by Eq. (\ref{Ga_inout}) at $g_{l}=2\,\textrm{MHz}$ (thick line) and $g_{l}=1\,\textrm{MHz}$ (thin line). Dashed curves are by Eq. (\ref{Ga_circuit}) at $g_{l}=2\,\textrm{MHz}$ (thick line) and $g_{l}=1\,\textrm{MHz}$ (thin line).}
\label{fig3} 
\end{figure}

Given the circuits in Fig.~\ref{fig1}, a semiclassical circuit theory can be applied to explain the cooling process. Let the voltage on the phase island (labeled as $\varphi$) be $v_{b}=\dot{\varphi}$. With a driving voltage $v_{c}e^{i\omega_{d}t}$ and the nanomechanical vibration $x(t)=xe^{i\omega_{a}t}$, we derive that
\begin{equation}
v_b=\frac{-(\omega_d+\omega_a)^2C_{x0}xv_c}{C_{\Sigma 0}d_0(\omega_b^2-(\omega_d+\omega_a)^2+i\kappa_0(\omega_d+\omega_a))}
\end{equation}
where the damping in the circuit is $\kappa_{0}=(RC_{\Sigma0})^{-1}$.  The dynamical backaction force on the nanomechanical resonator from the electromagnetic field can be written as
\begin{equation}
F_{e}=-\frac{\epsilon_{0}S_{0}}{2(d_{0}+x)^{2}}(v_{c}-v_{b})^{2}\approx \lambda x-m\Gamma_{c} \dot{x}
\end{equation}
which contains a small modification $\lambda$ to the elastic constant of the nanomechanical mode and a frictional force proportional to $\dot{x}$. Here, $S_{0}$ is the area of the capacitor plate of $C_{x}$. The frictional force has a $\pi/2$-phase difference with the nanomechanical motion with $\dot{x}\approx i\omega_{a}x$ at weak coupling.  It can be derived that
\begin{equation}
\Gamma_{c}=\frac{4g_{l}^{2}(\omega_d+\omega_a)^{3}\kappa_{0}/\omega_{b}}{((\omega_d+\omega_a)^{2}-\omega_{b}^{2})^{2}+(\omega_d+\omega_a)^{2}\kappa_{0}^{2}},\label{Ga_circuit}
\end{equation}
when substituting Eq.~(\ref{gl}) for $g_{l}$. At $-\Delta=\omega_{a}$, we have $\Gamma_{c}=4g_{l}^{2}/\kappa_{0}$. In Fig.~\ref{fig3}, we plot the cooling rates by Eq. (\ref{Ga_inout}) and by Eq. (\ref{Ga_circuit}). In the resolved-sideband regime with $\kappa_0\ll\omega_a$, the two results are only slightly different. However, the semiclassical approach can not explain the quantum backaction noise in Eq. (\ref{naf_inout}).

The cooling rate by the linear coupling in circuit (b) is proportional to $v_{c}^{2}$.  For comparison, the cooling rate by the radiation pressure-like coupling in circuit (a) is proportional to the average occupation number $\bar{n}_b$ determined by the driving voltage $v_g$ on the LC oscillator. With like parameters, it requires that $\bar{n}_b\approx 10^{6}$ in circuit (a) to achieve cooling rate comparable to that in the linear coupling scheme. 

To conclude, we studied a novel scheme for the cooling of nanomechanical resonators via linear coupling in a solid-state circuit. We showed that ground state cooling can be achieved in the resolved-sideband regime by periodically modulating the linear coupling. At moderate Q-factor, the final occupation number of the nanomechanical mode is lower bounded by the quantum backaction noise due to the counter rotating term in the linear coupling. The scheme can be realized in superconducting circuits with current technology and provides an interesting alternative to cooling schemes based on radiation pressure-like force.  

The author is grateful to Prof. Mark Kasevich and Prof. Dan M. Stamper-Kurn for stimulating discussions and to the Karel Urbanek Fellowship in the Dept. of Applied Physics at Stanford University for support.


\begin{thebibliography}{10}
\bibitem{SchwabPToday2005} K. C. Schwab and M. L. Roukes, Phys. Today
\textbf{58}, 36 (2005).

\bibitem{HarrisNature2008} J. D. Thompson and \emph{et al.}, Nature
\textbf{452}, 72 (2008).

\bibitem{Kippenberg2008}
 A. Schliesser and \emph{et al.} Nature Phys. \textbf{4}, 415 (2008).

\bibitem{RegalNPhys2008} C. A. Regal, J. D. Teufel, K. W. Lehnert, Nature Phys. \textbf{4}, 555 (2008); J. D. Teufel, J. W. Harlow, C. A. Regal, K. W. Lehnert, preprint
arXiv:0807.3585. 

\bibitem{BraginskyScience1980} V. B. Braginsky, Y. I. Vorontsov,
and K. S. Thorne, Science \textbf{209}, 547 (1980).

\bibitem{LeggettPRL1985} A. J. Leggett and A. Garg, Phys. Rev. Lett.
\textbf{54}, 857 (1985).

\bibitem{LFWeiPRL2006} L. F. Wei, Y. X. Liu, C. P. Sun, and F. Nori,
Phys. Rev. Lett. \textbf{97}, 237201 (2006).

\bibitem{MarshallPRL2003}
W. Marshall, C. Simon, R. Penrose, and D. Bouwmeester, Phys. Rev. Lett. \textbf{91}, 130401 (2003).

\bibitem{BlencowePRep2004} M. Blencowe, Phys. Rep. \textbf{395}, 159 (2004); A. D. Armour, M. P. Blencowe, and K. C. Schwab, Phys. Rev. Lett. \textbf{88}, 148301 (2002); A. N. Cleland and M. R. Geller, Phys. Rev. Lett. \textbf{93}, 070501 (2004);  M. Sarovar, H.-S. Goan, T. P. Spiller, and G. J. Milburn, Phys. Rev. A \textbf{72}, 062327 (2005).

\bibitem{TianZollerPRL2004} L. Tian and P. Zoller, Phys. Rev. Lett. \textbf{93}, 266403 (2004);  D. Stickand \emph{et al.}, Nature Phys. \textbf{2}, 36 (2006); P.
Treutlein and \emph{et al.}, Phys. Rev. Lett. \textbf{99}, 140403
(2007); S. Gupta, K. L. Moore, K. W. Murch, and D. M. Stamper-Kurn,
Phys. Rev. Lett. \textbf{99}, 213601 (2007).

\bibitem{JacobsPRL2007_2} K. Jacobs, Phys. Rev. Lett. \textbf{99},
117203 (2007);  H. Yuan and S. Lloyd, Phys. Rev. A \textbf{75}, 052331 (2007); 
L. Tian, Phys. Rev. B  \textbf{72}, 195411 (2005); D. H. Santamore, A. C. Doherty, and M. C. Cross, Phys. Rev. B \textbf{70}, 144301 (2004); I. Martin and W. H. Zurek, Phys. Rev. Lett. \textbf{98}, 120401 (2007); A. A. Clerk and D. W. Utami, Phys. Rev. A \textbf{75}, 042302 (2007).

\bibitem{KnobelNature2003} R. G. Knobel and A. N. Cleland, Nature
\textbf{424}, 291 (2003).

\bibitem{SchwabScience2006} M. D. LaHaye, O. Buu, B. Camarota, and K.
C. Schwab, Science \textbf{304}, 74 (2004); A. Naik and \emph{et al.}, Nature \textbf{443}, 193 (2006).

\bibitem{FlowersJacobsPRL2007}
N. E. Flowers-Jacobs, D. R. Schmidt and K. W. Lehnert, Phys. Rev. Lett. \textbf{98}, 
096804 (2007).

\bibitem{BrownPRL2007} K. R. Brown and \emph{et al.}, Phys. Rev.
Lett. \textbf{99}, 137205 (2007).

\bibitem{Karrai2004} C. H. Metzger and K. Karrai, Nature \textbf{432}, 1002 (2004); S. Gigan and \emph{et al.}, Nature \textbf{444}, 67 (2006); O. Arcizet and \emph{et al.},
Nature \textbf{444}, 71 (2006); A. Schliesser and \emph{et al.}, Phys.
Rev. Lett. \textbf{97}, 243905 (2006); T. Corbitt and \emph{et
al.}, Phys. Rev. Lett. \textbf{98}, 150802 (2007).

\bibitem{WilsonRaePRL2007} I. Wilson-Rae, N. Nooshi, W. Zwerger,
and T. J. Kippenberg, Phys. Rev. Lett. \textbf{99}, 093901 (2007).

\bibitem{GirvinPRL2007} F. Marquardt, J. P. Chen, A. A. Clerk, and
S. M. Girvin, Phys. Rev. Lett. \textbf{99}, 093902 (2007).

\bibitem{GenesPRA2008} C. Genes and \emph{et al.}, Phys. Rev. A \textbf{77},
033804 (2008).

\bibitem{ClerkPRL2006} A. A. Clerk, Phys. Rev. Lett. \textbf{96},
056801 (2006).

\bibitem{MartinPRB2004} I. Martin, A. Shnirman, L. Tian, and P. Zoller,
Phys. Rev. B \textbf{69}, 125339 (2004).

\bibitem{WilsonRaePRL2004} I. Wilson-Rae, P. Zoller, and A. Imamo$\bar{\textrm{g}}$lu, Phys. Rev. Lett. \textbf{92}, 075507 (2004).

\bibitem{TianResonators} L. Tian and S. M. Carr, \prb \textbf{74}, 125314 (2006).

\bibitem{Grifoni2004} M. Thorwart, E. Paladino, and M. Grifoni, Chem.
Phy. \textbf{296}, 333 (2004); F. K. Wilhelm, S. Kleff, and J. von
Delft, Chem. Phys. \textbf{296}, 345 (2004).

\bibitem{WallsBook1994} D. F. Walls and G. J. Milburn, \emph{Quantum Optics},
(Springer, Berlin, 1994).

\bibitem{KippenbergPreprint} Upon finishing this work, we found the preprint arXiv:0805.2528 by J.M. Dobrindt, I. Wilson-Rae, and T.J. Kippenberg, which addressed the origin of quantum backaction noise in radiation pressure cooling.

\bibitem{WinelandPRA1987} D. J. Wineland, W. M. Itano, J. C. Bergquist,
and R. G. Hulet, Phys. Rev. A \textbf{36}, 2220 (1987). 

\bibitem{MakhlinRMP2001} Y. Makhlin, G. Sch\"{o}n and A. Shnirman, Rev.
Mod. Phys. \textbf{73}, 357 (2001).

\bibitem{HouckNature2007} A. A. Houck and \emph{et al.}, Nature \textbf{449},
328 (2007); J. Majer and \emph{et al.}, Nature \textbf{449}, 443 (2007).

\bibitem{LarsonPRL1986} D. J. Larson and \emph{et al.}, Phys. Rev.
Lett. \textbf{57}, 70 (1986); C. J. Myatt and \emph{et al.}, Phys.
Rev. Lett. \textbf{78}, 586 (1997).

\end{thebibliography}
\end{document}